\documentclass[aps,prl,twocolumn,superscriptaddress,amsmath,amssymb]{revtex4}
\usepackage{graphicx,overpic} 
\usepackage{epstopdf} 
\usepackage{dcolumn} 
\usepackage{xcolor} 
\usepackage{hyperref} 
\usepackage[utf8]{inputenc} 
\usepackage[english]{babel} 
\usepackage[displaymath, mathlines]{lineno} 
\usepackage{blindtext} 
\usepackage{amsmath} 
\usepackage{amssymb}
\usepackage{amsfonts}
\usepackage{upgreek} 
\usepackage{booktabs}
\usepackage[normalem]{ulem}
\usepackage{mathrsfs}
\usepackage{orcidlink}

\newcommand*\patchAmsMathEnvironmentForLineno[1]{%
\expandafter\let\csname old#1\expandafter\endcsname\csname #1\endcsname
\expandafter\let\csname oldend#1\expandafter\endcsname\csname
end#1\endcsname
 \renewenvironment{#1}%
   {\linenomath\csname old#1\endcsname}%
   {\csname oldend#1\endcsname\endlinenomath}%
}
\newcommand*\patchBothAmsMathEnvironmentsForLineno[1]{%
  \patchAmsMathEnvironmentForLineno{#1}%
  \patchAmsMathEnvironmentForLineno{#1*}%
}
\AtBeginDocument{%
\patchBothAmsMathEnvironmentsForLineno{equation}%
\patchBothAmsMathEnvironmentsForLineno{align}%
\patchBothAmsMathEnvironmentsForLineno{flalign}%
\patchBothAmsMathEnvironmentsForLineno{alignat}%
\patchBothAmsMathEnvironmentsForLineno{gather}%
\patchBothAmsMathEnvironmentsForLineno{multline}%
\patchBothAmsMathEnvironmentsForLineno{eqnarray}%
}

\graphicspath{{figures/}} 

\usepackage{relsize}
\def\babar{\mbox{\slshape B\kern-0.1em{\smaller A}\kern-0.1em
    B\kern-0.1em{\smaller A\kern-0.2em R}}\xspace}

\newcommand{\lumi}{\ensuremath{207.2\invfb}\xspace}

\newcommand{\Lc}{\ensuremath{\Lambda^{+}_c}\xspace}
\newcommand{\p}{\ensuremath{p}\xspace}
\newcommand{\LcTopKpi}{\ensuremath{\Lc\to\p\Km\pip}\xspace}

\newcommand{\sigmat}{\ensuremath{\sigma_t}\xspace}

\newcommand{\tauLc}{203.20}
\newcommand{\tauLcStat}{0.89}
\newcommand{\tauLcSyst}{0.77}
\input{belle2sym}

\usepackage{ulem}

\begin{document}

\title{Measurement of the \Lc Lifetime}
  \author{F.~Abudin{\'e}n\,\orcidlink{0000-0002-6737-3528}} 
  \author{L.~Aggarwal\,\orcidlink{0000-0002-0909-7537}} 
  \author{H.~Ahmed\,\orcidlink{0000-0003-3976-7498}} 
  \author{J.~K.~Ahn\,\orcidlink{0000-0002-5795-2243}} 
  \author{H.~Aihara\,\orcidlink{0000-0002-1907-5964}} 
  \author{N.~Akopov\,\orcidlink{0000-0002-4425-2096}} 
  \author{A.~Aloisio\,\orcidlink{0000-0002-3883-6693}} 
  \author{N.~Anh~Ky\,\orcidlink{0000-0003-0471-197X}} 
  \author{D.~M.~Asner\,\orcidlink{0000-0002-1586-5790}} 
  \author{H.~Atmacan\,\orcidlink{0000-0003-2435-501X}} 
% ...
\collaboration{The Belle II Collaboration}

\begin{abstract}
An absolute measurement of the \Lc lifetime is reported using \LcTopKpi decays in events reconstructed from data collected by the \belletwo experiment at the SuperKEKB asymmetric-energy electron-positron collider. The total integrated luminosity of the data sample, which was collected at center-of-mass energies at or near the $\Upsilon(4S)$ resonance, is \lumi. The result, $\tau(\Lc) = \tauLc\pm\tauLcStat\pm\tauLcSyst\fs$, where the first uncertainty is statistical and the second systematic, is the most precise measurement to date and is consistent with previous determinations.
\end{abstract}

\maketitle

Searches for physics beyond the standard model of particle physics through precise measurements of 
weakly decaying charm or bottom hadrons often rely on accurate theoretical descriptions of strong 
interactions at low energy, typically using effective models such as the heavy quark expansion 
(HQE)~\cite{Neubert:1997gu,Uraltsev:2000qw,Lenz:2013aua,Lenz:2014jha,Kirk:2017juj,Cheng:2018rkz,Gratrex:2022}.
The HQE provides a consistent framework for computing the decay widths of heavy hadrons in terms of
inverse powers of the heavy quark mass. For bottom hadrons, nonperturbative effects are relatively 
small and the HQE in 1/$m_{b}$, where $m_{b}$ is the mass of the bottom quark, works well. 
In contrast, higher-order corrections due to the influence of light valence (spectator) quarks 
are significant for charm hadron lifetimes, for which the HQE to 1/$m_{c}^3$ does not satisfactorily describe 
lifetimes~\cite{Gratrex:2022}. The lifetimes of the $\Omega_c^0$ and $\Xi_c^0$ were recently measured
to be much larger than the previous world average~\cite{Aaij:2018dso,Aaij:2019lwg,Aaij:2021}, inverting 
the known hierarchy of charm lifetimes. Careful consideration of model-dependent spectator effects
is required for theoretical predictions of charm baryon lifetimes to agree with experimental 
measurements~\cite{Cheng:2018rkz,Gratrex:2022}. Improved measurements of charm baryon lifetimes 
therefore provide refined tests for effective models.

The world average value of the \Lc lifetime is $202.4\pm3.1\fs$~\cite{pdg}. Previous measurements include
percent-level results from the FOCUS, SELEX, and CLEO collaborations two decades 
ago~\cite{Link:2002uhq,Kushnirenko:2001,CLEO:2001}, as well as a more precise measurement, relative 
to the \Dp lifetime, from the LHCb collaboration~\cite{Aaij:2019lwg}. The latter of these has a limiting systematic 
uncertainty associated with the \Dp lifetime. Relative measurements minimize systematic uncertainties 
related to event selection that may bias the decay time, particularly at hadron colliders. In contrast, the ability 
to reconstruct charm hadrons without biasing the decay time allows experiments at electron-positron (\epem) colliders to precisely 
determine absolute lifetimes, as demonstrated by the recent measurement of the \Dz and \Dp 
lifetimes~\cite{Abudinen:2021} from the \belletwo experiment~\cite{Abe:2010gxa} at the SuperKEKB 
asymmetric-energy \epem collider~\cite{Akai:2018mbz}. The most recent \Lc lifetime measurement at an 
\epem collider, from the CLEO collaboration, is in mild tension with other results and increases the 
uncertainty of the world average~\cite{pdg}. A precise, absolute measurement by Belle~II may help to resolve
the tension between \Lc lifetime measurements at \epem colliders and other experiments and will substantially improve 
the world average.

In this Letter, we report a precise measurement of the \Lc lifetime using $\Lc\to\p\Km\pip$ decays reconstructed in 
data collected at or near the $\Upsilon(4S)$ resonance, corresponding to a center-of-mass energy at or near 10.58 GeV, by the \belletwo experiment from 2019 to 
mid 2021 and corresponding to an integrated luminosity of \lumi. Unless specified otherwise, charge conjugate 
decays are implied throughout. 

The lifetime of the \Lc is determined from a two-dimensional fit to the decay time $t$ and its uncertainty \sigmat.
The decay time is calculated assuming that \Lc candidates are promptly produced from
continuum $\epem\rightarrow c\bar{c}$ events and is determined according to 
$t = m_{\Lambda_{c}}\vec{L}\cdot\vec{p}/|\vec{p}|^2$, where $m_{\Lambda_{c}}$ is the world average mass of the 
\Lc~\cite{pdg}, $\vec{L}$ is the displacement of the \Lc decay point from the \epem interaction point (IP), 
and $\vec{p}$ is the momentum of the \Lc candidate. The position and size of the IP region is determined using
$e^{+}e^{-}\rightarrow\mu^{+}\mu^{-}$ events. Event selection criteria and the fit strategy are optimized 
and validated using simulated data, but the fit to the collision data does not use any input taken from simulation. 

The \belletwo detector~\cite{Abe:2010gxa} includes a tracking system comprising a two-layer silicon pixel detector 
(PXD) surrounded by a four-layer double-sided silicon strip detector (SVD) and a 56-layer central drift chamber (CDC). 
The second layer of the PXD had 15\% azimuthal coverage during the collection of the data used in this study. For 
the \Lc decays considered here, the combined PXD and SVD vertexing system provide a decay-length resolution of 
40\mum, corresponding to an average decay-time resolution of 87\fs for an average decay length of 96\mum. A 
time-of-propagation counter in the barrel region of the detector and an aerogel ring-imaging Cherenkov counter in the 
endcap region provide charged-particle identification (PID) information. An electromagnetic calorimeter consisting of 
CsI(Tl) crystals provides energy and timing measurements for photons and electrons. A $K^0_L$ and muon detector is 
installed in the iron flux return yoke of a superconducting solenoid magnet that provides a $1.5\,\rm{T}$ magnetic field.

We generate $e^+e^-\rightarrow q\bar{q}$ events with \texttt{KKMC}~\cite{Jadach:1999vf} and hadronize quarks 
with \texttt{Pythia~8}~\cite{Sjostrand:2014zea}. Hadron decays are emulated using \texttt{EvtGen}~\cite{Lange:2001uf}. 
The detector response is simulated with \texttt{Geant4}~\cite{Agostinelli:2002hh}. Reconstruction of events from 
simulated and collision data is performed with the Belle~II analysis software framework~\cite{Kuhr:2018lps}. In 
addition to the excellent vertexing capability, Belle~II benefits from good charged-particle tracking 
performance~\cite{BelleIITrackingGroup:2020hpx,Simon:1960}.

Candidate \LcTopKpi decays are each reconstructed from one negatively and two positively charged particles, which
are required to be well measured with reliable uncertainties to allow for a precise lifetime measurement. Each charged 
particle must be associated with one or more PXD measurements (hits) in the PXD, at least one hit in the first layer of 
the SVD, and at least 20 hits in the CDC. Each charged particle must have a distance of closest approach to the IP of 
less than 0.5 cm in the plane transverse to the beam and 2 cm in the direction parallel to it to remove charged particles 
not associated with the \epem interaction. A fit constrains the charged particles to come from a common vertex and the 
\Lc candidate to come from the IP~\cite{treefitter}. Candidates with a vertex-fit $\chi^2$ probability less than 0.01 are 
rejected. Since the \Lc is assumed to originate from the IP, secondary decays in which the \Lc originates from a displaced 
vertex would bias the lifetime measurement. To suppress \Lc from $B$ decays, the center-of-mass momentum of each 
\Lc candidate is required to be greater than 2.5 GeV/$c$.

Charged PID information is combined from all subdetector systems except the PXD and SVD. This PID information 
is used to construct likelihoods $\mathscr{L}(h)$ for a given particle hypothesis $h$. For each candidate, one 
positively charged particle is required to have $\mathscr{L}(p)/(\mathscr{L}(p)+\mathscr{L}(K)+\mathscr{L}(\pi)) > 0.9$, 
the negatively charged particle is required to have 
$\mathscr{L}(K)/(\mathscr{L}(p)+\mathscr{L}(K)+\mathscr{L}(\pi)+\mathscr{L}(\mu)+\mathscr{L}(e)+\mathscr{L}(d)) > 0.6$, 
and the remaining positively charged particle is assumed to be a pion. Here $\mathscr{L}(d)$ is the deuteron hypothesis likelihood.
The efficiency of the proton identification is found to 
be about 88\%, with a kaon contamination of less than 2\%, and the efficiency of kaon identification is 70\%, with a pion 
contamination of 6\%, from studies of $\Lambda^{0}\rightarrow p\pi^{-}$ and $D^{*+}$-tagged $D^{0}\rightarrow K^{-}\pi^{+}$ 
decays. To reduce backgrounds from misidentified charm-meson decays, we reject events with $M(\pi^{+}K^{-}\pi^{+})$ in 
[1.858, 1.881] GeV/$c^2$ or [2.000, 2.020] GeV/$c^2$, with both positively charged particles assumed to be pions. These 
intervals correspond to three units of resolution, or standard deviations, on the reconstructed mass in both directions around 
the known $D^{+}$ and $D^{*+}$ masses, respectively. Other charm-related backgrounds are suppressed by requiring that 
the transverse momenta of pions exceed 0.35 GeV/$c$ and those of protons exceed 0.7 GeV/$c$. 

Events with multiple candidates, which occur at a rate of 0.2\%, are rejected. Analysis of simulated 
events shows that the event selection criteria do not bias the measurement of the \Lc lifetime.

Decays of $\Xi_c^{0}\rightarrow\pi^{-}\Lambda_c^+$ and $\Xi_c^{+}\rightarrow\pi^{0}\Lambda_c^+$ may bias
the measurement of the \Lc lifetime, since the $\Xi_c^0$ and $\Xi_c^+$ have lifetimes of
$153\pm 6\fs$ and $456\pm 5\fs$~\cite{pdg}, respectively, and may shift the production vertex of the 
\Lc away from the IP. The amount of $\Xi_c$ contamination is estimated from a fit to the distribution of the 
\Lc vertex displacement from the IP in the plane transverse to the beam line. This distribution depends
only on the resolution of the detector for \Lc candidates that are produced at the IP. \Lc candidates
from $\Xi_c$ decays have production vertices that are displaced from the IP and therefore a larger vertex displacement from the IP. The fit to the distribution of the \Lc transverse vertex displacement
gives a background contamination of 374 $\pm$ 88 events, corresponding to 0.003\% of \Lc candidates. As this includes both combinatoric backgrounds and $\Xi_c$ decays,
the central value is taken as an estimate of the maximum number of $\Xi_c$ decays. This value is consistent with
predictions based on the expected production cross-sections for $\Xi_{c}^{0}$ and $\Xi_{c}^{+}$~\cite{bxic},
the measured branching fraction for $\Xi_c^{0} \rightarrow \pi^{-}\Lambda_c^+$~\cite{xiclhcb},
and theoretical predictions for $\Xi_c^{+} \rightarrow \pi^{0}\Lambda_c^+$~\cite{xicppred}. 
Backgrounds from $\Xi_c$ decays are reduced by restricting the invariant
mass of the $\Xi_c$ candidate formed by combining the \Lc candidate with a $\pi^{-}$ or $\pi^{0}$ from
the unassigned particle candidates of the event. This restriction is optimized using simulations and the optimal precision
on the lifetime measurement is achieved by restricting the mass difference between the $\Xi_c$ and \Lc candidates to
two units of the mass resolution, removing events with $M(pK^{-}\pi^{+}\pi^{-})$-$M(pK^{-}\pi^{+})$ in [0.1834, 0.1864] GeV/$c^2$ or
$M(pK^{-}\pi^{+}\pi^{0})$ - $M(pK^{-}\pi^{+})$ in [0.1753, 0.1873] GeV/$c^2$. About 61\% of $\Xi_{c}$ decays to $\Lambda_{c}$
survive this veto, according to studies of simulated events. To account for the effect of the remaining 
background of this type, the measured \Lc lifetime is corrected by subtracting a bias of 0.34 fs, as discussed below.

\begin{figure}[t!]
\centering
\includegraphics[width=\linewidth]{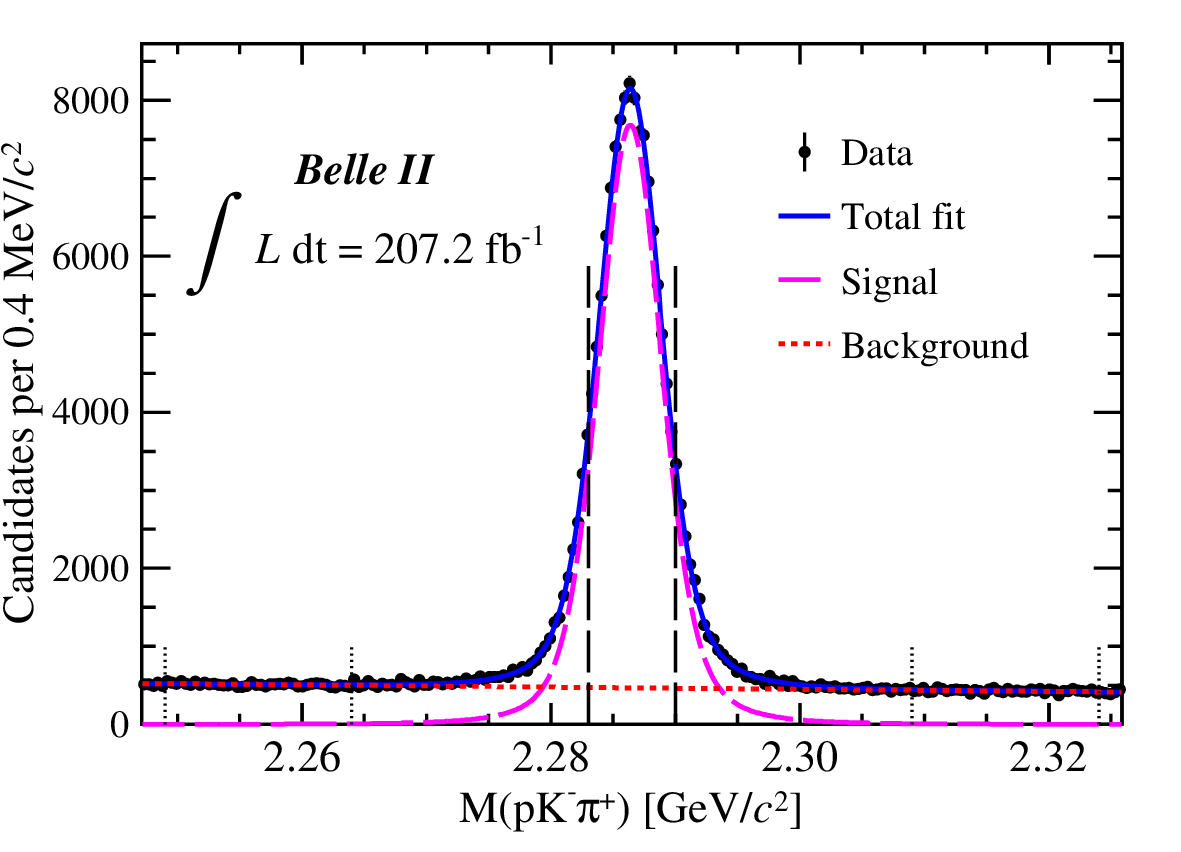}\\
\caption{Mass distribution of \LcTopKpi candidates with fit projections overlaid. The vertical dashed lines 
enclose the signal region and the short, vertical dotted lines enclose the sidebands.\label{fig:massfit}}
\end{figure}

After all selection criteria, the number of events in the \Lc signal range, defined as 
$M(pK^{-}\pi^{+})$ in [2.283, 2.290] GeV/$c^{2}$, within about 1.4 units of the mass resolution
around the world average \Lc mass, is $1.16\times10^5$. The relative amount of signal events, determined 
from a binned least-squares fit to the $M(pK^{-}\pi^{+})$ distribution (Fig.~\ref{fig:massfit}), is 92.5\%. 
In the fit, the \LcTopKpi signal is modeled as the sum of Gaussian and Johnson functions~\cite{johnson} with 
a common mode. The background is modeled as a linear function. Events from the \Lc sidebands, defined as 
$M(pK^{-}\pi^{+})$ in [2.249, 2.264] GeV/$c^{2}$ and [2.309, 2.324] GeV/$c^{2}$, are used
to constrain the background in the lifetime fit.

The \Lc lifetime is measured with an unbinned maximum-likelihood fit to the $(t,\sigmat)$ distribution for 
events in the signal region. The signal probability density function (PDF) is the product of an exponential 
function in $t$ convolved with a Gaussian resolution function, which depends on \sigmat, and a PDF for \sigmat. 
The latter is a histogram template formed from signal candidates subtracted by the distribution of sideband candidates 
after scaling according to the size of the signal and background regions. To account for a possible bias in the 
decay-time determination, the mean of the resolution function is determined by the fit.

The background PDF, an empirical model of the sideband data, is the sum of two exponential functions
convolved with Gaussian resolution functions, which account for backgrounds from long-lived particles, 
and a zero-lifetime component consisting only of the resolution function, which accounts for combinatorial backgrounds. 
To account for a possible misestimation of the decay-time uncertainty, the width of the resolution function is given by the 
per-candidate \sigmat multiplied by a scale factor $s$, which is a free parameter in the lifetime fit. The mean of the 
resolution function is common for all terms, but a separate \sigmat-scaling parameter is used for the background PDF.  

To better constrain the background, a simultaneous fit to the events in the signal region and sidebands is performed, 
where the \sigmat PDF for the sidebands is a binned template determined by sideband events. The background fraction in 
the lifetime fit is Gaussian constrained to $(7.50 \pm 0.02)\%$, as determined from the $M(pK^{-}\pi^{+})$ fit.

The lifetime fit is validated both on fully simulated data equivalent to 1 ab$^{-1}$, about five times the integrated 
luminosity of the collision data, and on simulated distributions generated by randomly sampling the lifetime PDF 
determined from a fit to the collision data. All validation fits return unbiased results, regardless of 
the assumed \Lc lifetime. Studies of the decay-time distribution in simulation suggest that \sigmat is underestimated 
by about 10\%, which is in good agreement with the results from the lifetime fit to the data, for which the scale parameter
is determined to be $s$ = 1.108 $\pm$ 0.006. The mean of the resolution function is determined to be 4.77 $\pm$ 0.63 fs.

\begin{figure}[t!]
\centering
\includegraphics[width=\linewidth]{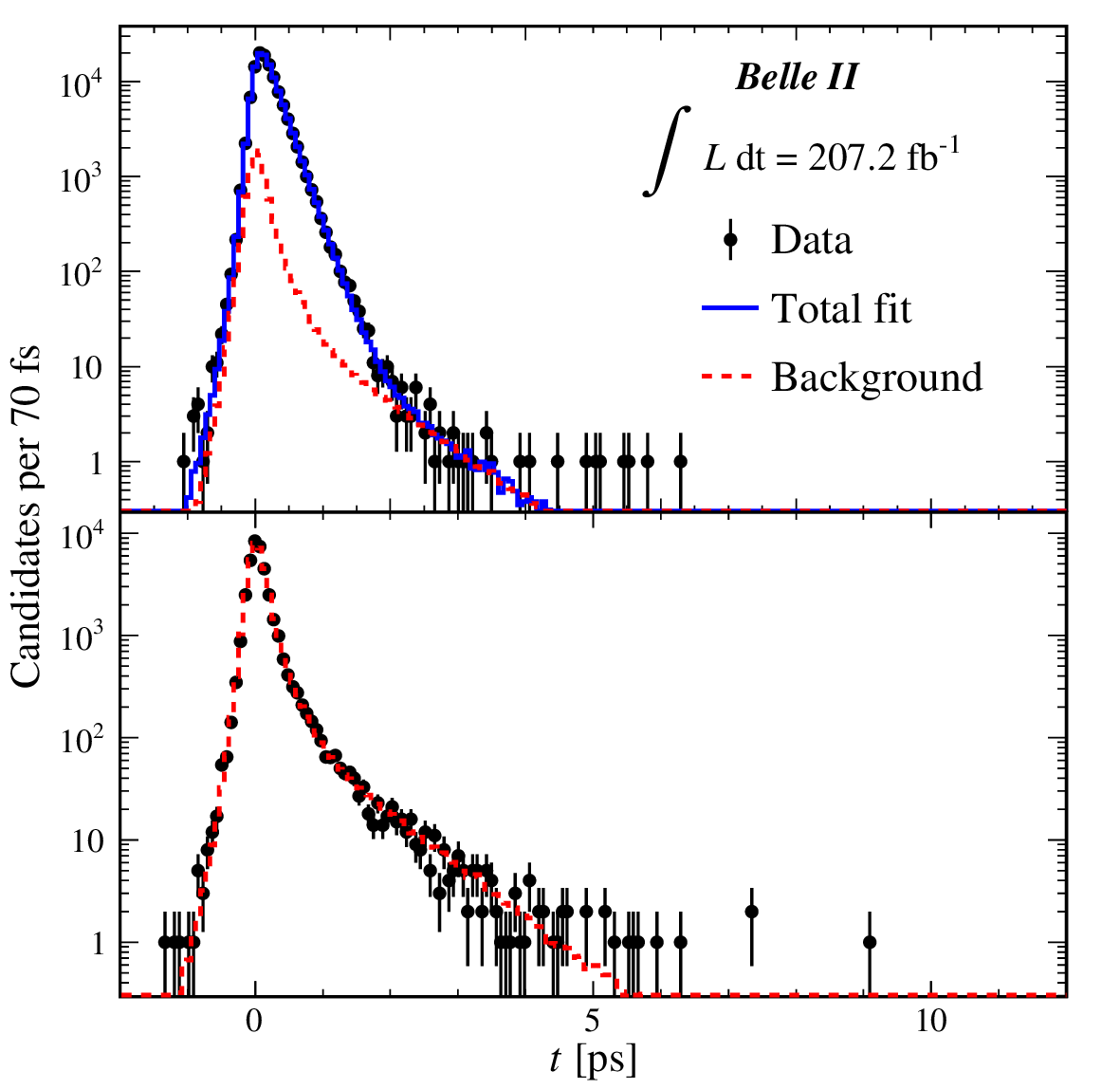}\\
\caption{Decay-time distribution of \LcTopKpi events in the signal region (top) and sidebands (bottom) with fit 
projections overlaid.
\label{fig:lifetime-fit}}
\end{figure}

\begin{figure}[t!]
\centering
\includegraphics[width=\linewidth]{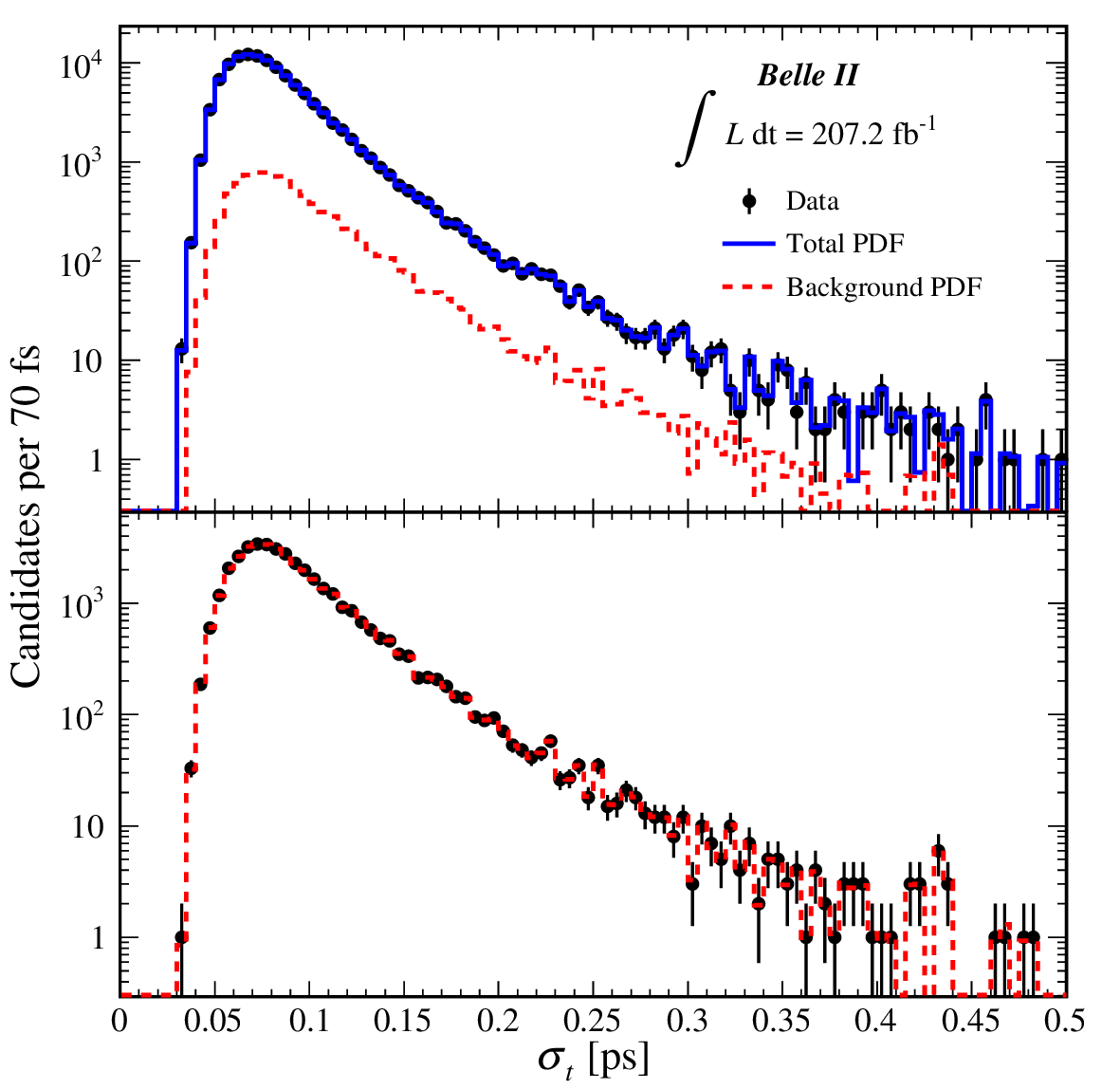}\\
\caption{Decay-time uncertainty distribution of \LcTopKpi events in the signal region (top) and sidebands (bottom).
The \sigmat PDF used in the fit is shown by the solid blue histogram and the background \sigmat PDF is shown by the 
dashed red histogram. \label{fig:lifetime-sigmat}}
\end{figure}

The \Lc lifetime is measured to be $\tauLc\pm\tauLcStat\fs$, where the uncertainty is statistical only. 
The lifetime fit projection, overlaid on the decay-time distribution in the data sample, 
is shown in Fig.~\ref{fig:lifetime-fit}. The \sigmat PDF used in the lifetime fit is shown in Fig.~\ref{fig:lifetime-sigmat}.
The systematic uncertainty is calculated from the sum in quadrature of individual contributions from the sources 
listed in Table~\ref{tab:syst} and described below.

\begin{table}[t]
\centering
\caption{Systematic uncertainties on the \Lc lifetime\label{tab:syst}}
\begin{tabular}{lcc}
\hline\hline
Source & Uncertainty [fs] \\
\hline
$\Xi_{c}$ contamination & 0.34 \\
Resolution model   & 0.46 \\
Non-$\Xi_{c}$ backgrounds & 0.20 \\
Detector alignment & 0.46 \\
Momentum scale     & 0.09 \\
\hline
Total  & \tauLcSyst \\
\hline\hline
\end{tabular}
\end{table}

The systematic uncertainty due to backgrounds from $\Xi_c$ decays is determined by adding
simulated events of this type to the 1~ab$^{-1}$ equivalent simulated sample according to
the estimated maximum contamination determined from the fit to the distribution 
of the \Lc transverse vertex displacement in data and repeating the measurement. The difference between the 
simulated \Lc lifetime and the measured value is 0.68 fs. Since this is an estimate of the
maximum effect of remaining $\Xi_c$ backgrounds, half the difference, 0.34 fs, is taken as both a 
correction to the lifetime and an associated systematic uncertainty.

The resolution model for the lifetime PDF is complicated by correlations between the decay time and the 
decay-time uncertainty such that it cannot be described by a simple Gaussian function. We neglect these 
correlations in our model, which consists of a \sigmat-dependent Gaussian resolution multiplied by a PDF 
in \sigmat, and include the impact of this approximation as a systematic uncertainty. We fit our model to 
1000 sets of signal-only simulated decays, each with a size equivalent to the data. The sets are produced 
by resampling, with repetition, simulated events in an amount corresponding to an equivalent luminosity of
1~ab$^{-1}$. The difference in the mean lifetime determined from these fits relative to the true value is $0.46\fs$, 
which is taken as a systematic uncertainty due to the resolution model. 

To check the resolution model, the lifetime fit is repeated with the Gaussian resolution function replaced with a 
sum of two Gaussian functions. The difference in the measured lifetime, $0.36\pm0.23\fs$, is covered by the 
corresponding systematic uncertainty. The bias of the decay-time resolution function for signal events 
depends on the \Lc candidate mass, but cancels if the signal range is centered on the true mass. 
Differences in the measured lifetime with the signal region varied are consistent with statistical fluctuations 
and are within the systematic uncertainty due to the resolution model.

Sideband events are included in the lifetime fit to constrain the background PDF. In simulation, sideband 
events describe the background distribution in the signal region accurately. To account 
for potential disagreements between the signal region and sidebands in the data, we produce 1000 sets of
simulated data by resampling from the 1~ab$^{-1}$-equivalent simulated sample for
events in the signal region and from the sidebands of the data sample for events in the sideband region. 
The mean lifetime residual is $0.20\fs$, which is taken as a
systematic uncertainty associated with background contamination. 

To check the signal PDF for the $M(pK^{-}\pi^{+})$ fit, we replace the sum of Gaussian and Johnson functions with a sum of two
Gaussian functions. Using the resulting background contribution has a negligible effect on the measured lifetime.

Reconstruction of charged particles at \belletwo relies on periodic calibrations to correct for detector misalignment 
and surface deformations of the internal components of the PXD and SVD, as well as for relative alignments of the 
tracking system~\cite{Bilka:2020kgr}. Detector misalignment can bias measured particle-decay lengths and therefore 
their decay times. To account for imperfections in the detector alignment, sets of signal-only simulated data, each with a size 
comparable to the collision data, are produced with detectors randomly misaligned according  to the alignment precision 
observed in data. The root mean square dispersion of the lifetime residuals in these misaligned simulated datasets is 
0.46 fs, which is taken as a systematic uncertainty due to imperfect detector alignment.

The momenta of charged particles are scaled by a factor, 0.99971, determined by calibrating the peak positions of 
abundant charm, strange, and bottom hadron decays. The uncertainty on this scale factor, 0.0009, results in 
a systematic uncertainty on the \Lc lifetime of $0.09\fs$. The uncertainty on the world average of 
the \Lc mass results in a negligible systematic uncertainty.

As a check of the internal consistency of the lifetime measurement, the full analysis is repeated on subsets 
of data chosen according to data-collection periods and \Lc momentum ranges, directions, and charge. The
result for each subset is consistent with the full result. The lifetime fit is also repeated by selecting the candidate 
with the best vertex fit probability or randomly selecting a candidate, rather than rejecting events with multiple candidates. 
The difference in lifetime in each case is negligible. Finally, several events in the data have lifetimes greater
than 4 ps, as shown in Fig.~\ref{fig:lifetime-fit}. Studies of simulated events suggest that these are from long-lived
charm meson decays and show that they do not bias the lifetime result with the current dataset size.

In conclusion, we measure the \Lc lifetime to be $\tauLc\pm\tauLcStat\pm\tauLcSyst\fs$, where the first uncertainty is
statistical and the second systematic, using
data with an integrated luminosity of \lumi collected by the \belletwo experiment at the SuperKEKB 
asymmetric-energy \epem collider. This is consistent with the recent, relative measurement by 
LHCb~\cite{Aaij:2019lwg} and other previous results, though the mild tension between the measurement 
by CLEO~\cite{CLEO:2001} and all other measurements remains. The absolute measurement presented here 
is the most precise \Lc lifetime measurement to date and may be useful to test the accuracy of HQE models as 
theoretical precision improves and discrepancies between theory and experiment become more significant.

\begin{acknowledgments}

This work, based on data collected using the Belle II detector, which was built and commissioned prior to March 2019, was supported by
Science Committee of the Republic of Armenia Grant No.~20TTCG-1C010;
Australian Research Council and research Grants
No.~DE220100462,
No.~DP180102629,
No.~DP170102389,
No.~DP170102204,
No.~DP150103061,
No.~FT130100303,
No.~FT130100018,
and
No.~FT120100745;
Austrian Federal Ministry of Education, Science and Research,
Austrian Science Fund
No.~P~31361-N36
and
No.~J4625-N,
and
Horizon 2020 ERC Starting Grant No.~947006 ``InterLeptons'';
Natural Sciences and Engineering Research Council of Canada, Compute Canada and CANARIE;
Chinese Academy of Sciences and research Grant No.~QYZDJ-SSW-SLH011,
National Natural Science Foundation of China and research Grants
No.~11521505,
No.~11575017,
No.~11675166,
No.~11761141009,
No.~11705209,
and
No.~11975076,
LiaoNing Revitalization Talents Program under Contract No.~XLYC1807135,
Shanghai Pujiang Program under Grant No.~18PJ1401000,
and the CAS Center for Excellence in Particle Physics (CCEPP);
the Ministry of Education, Youth, and Sports of the Czech Republic under Contract No.~LTT17020 and
Charles University Grant No.~SVV 260448 and
the Czech Science Foundation Grant No.~22-18469S;
European Research Council, Seventh Framework PIEF-GA-2013-622527,
Horizon 2020 ERC-Advanced Grants No.~267104 and No.~884719,
Horizon 2020 ERC-Consolidator Grant No.~819127,
Horizon 2020 Marie Sklodowska-Curie Grant Agreement No.~700525 "NIOBE"
and
No.~101026516,
and
Horizon 2020 Marie Sklodowska-Curie RISE project JENNIFER2 Grant Agreement No.~822070 (European grants);
L'Institut National de Physique Nucl\'{e}aire et de Physique des Particules (IN2P3) du CNRS (France);
BMBF, DFG, HGF, MPG, and AvH Foundation (Germany);
Department of Atomic Energy under Project Identification No.~RTI 4002 and Department of Science and Technology (India);
Israel Science Foundation Grant No.~2476/17,
U.S.-Israel Binational Science Foundation Grant No.~2016113, and
Israel Ministry of Science Grant No.~3-16543;
Istituto Nazionale di Fisica Nucleare and the research grants BELLE2;
Japan Society for the Promotion of Science, Grant-in-Aid for Scientific Research Grants
No.~16H03968,
No.~16H03993,
No.~16H06492,
No.~16K05323,
No.~17H01133,
No.~17H05405,
No.~18K03621,
No.~18H03710,
No.~18H05226,
No.~19H00682, 
No.~22H00144,
No.~26220706,
and
No.~26400255,
the National Institute of Informatics, and Science Information NETwork 5 (SINET5), 
and
the Ministry of Education, Culture, Sports, Science, and Technology (MEXT) of Japan;  
National Research Foundation (NRF) of Korea Grants
No.~2016R1\-D1A1B\-02012900,
No.~2018R1\-A2B\-3003643,
No.~2018R1\-A6A1A\-06024970,
No.~2018R1\-D1A1B\-07047294,
No.~2019R1\-I1A3A\-01058933,
No.~2022R1\-A2C\-1003993,
and
No.~RS-2022-00197659,
Radiation Science Research Institute,
Foreign Large-size Research Facility Application Supporting project,
the Global Science Experimental Data Hub Center of the Korea Institute of Science and Technology Information
and
KREONET/GLORIAD;
Universiti Malaya RU grant, Akademi Sains Malaysia, and Ministry of Education Malaysia;
Frontiers of Science Program Contracts
No.~FOINS-296,
No.~CB-221329,
No.~CB-236394,
No.~CB-254409,
and
No.~CB-180023, and No.~SEP-CINVESTAV research Grant No.~237 (Mexico);
the Polish Ministry of Science and Higher Education and the National Science Center;
the Ministry of Science and Higher Education of the Russian Federation,
Agreement No.~14.W03.31.0026, and
the HSE University Basic Research Program, Moscow;
University of Tabuk research Grants
No.~S-0256-1438 and No.~S-0280-1439 (Saudi Arabia);
Slovenian Research Agency and research Grants
No.~J1-9124
and
No.~P1-0135;
Agencia Estatal de Investigacion, Spain
Grant No.~RYC2020-029875-I
and
Generalitat Valenciana, Spain
Grant No.~CIDEGENT/2018/020
Ministry of Science and Technology and research Grants
No.~MOST106-2112-M-002-005-MY3
and
No.~MOST107-2119-M-002-035-MY3,
and the Ministry of Education (Taiwan);
Thailand Center of Excellence in Physics;
TUBITAK ULAKBIM (Turkey);
National Research Foundation of Ukraine, project No.~2020.02/0257,
and
Ministry of Education and Science of Ukraine;
the U.S. National Science Foundation and research Grants
No.~PHY-1913789 
and
No.~PHY-2111604, 
and the U.S. Department of Energy and research Awards
No.~DE-AC06-76RLO1830, 
No.~DE-SC0007983, 
No.~DE-SC0009824, 
No.~DE-SC0009973, 
No.~DE-SC0010007, 
No.~DE-SC0010073, 
No.~DE-SC0010118, 
No.~DE-SC0010504, 
No.~DE-SC0011784, 
No.~DE-SC0012704, 
No.~DE-SC0019230, 
No.~DE-SC0021274, 
No.~DE-SC0022350; 
and
the Vietnam Academy of Science and Technology (VAST) under Grant No.~DL0000.05/21-23.

These acknowledgements are not to be interpreted as an endorsement of any statement made
by any of our institutes, funding agencies, governments, or their representatives.

We thank the SuperKEKB team for delivering high-luminosity collisions;
the KEK cryogenics group for the efficient operation of the detector solenoid magnet;
the KEK computer group and the NII for on-site computing support and SINET6 network support;
and the raw-data centers at BNL, DESY, GridKa, IN2P3, INFN, and the University of Victoria for offsite computing support.
\end{acknowledgments}

\bibliographystyle{apsrev4-1}
\bibliography{references}

\begin{thebibliography}{30}%
\makeatletter
\providecommand \@ifxundefined [1]{%
 \@ifx{#1\undefined}
}%
\providecommand \@ifnum [1]{%
 \ifnum #1\expandafter \@firstoftwo
 \else \expandafter \@secondoftwo
 \fi
}%
\providecommand \@ifx [1]{%
 \ifx #1\expandafter \@firstoftwo
 \else \expandafter \@secondoftwo
 \fi
}%
\providecommand \natexlab [1]{#1}%
\providecommand \enquote  [1]{``#1''}%
\providecommand \bibnamefont  [1]{#1}%
\providecommand \bibfnamefont [1]{#1}%
\providecommand \citenamefont [1]{#1}%
\providecommand \href@noop [0]{\@secondoftwo}%
\providecommand \href [0]{\begingroup \@sanitize@url \@href}%
\providecommand \@href[1]{\@@startlink{#1}\@@href}%
\providecommand \@@href[1]{\endgroup#1\@@endlink}%
\providecommand \@sanitize@url [0]{\catcode `\\12\catcode `\$12\catcode
  `\&12\catcode `\#12\catcode `\^12\catcode `\_12\catcode `\%12\relax}%
\providecommand \@@startlink[1]{}%
\providecommand \@@endlink[0]{}%
\providecommand \url  [0]{\begingroup\@sanitize@url \@url }%
\providecommand \@url [1]{\endgroup\@href {#1}{\urlprefix }}%
\providecommand \urlprefix  [0]{URL }%
\providecommand \Eprint [0]{\href }%
\providecommand \doibase [0]{http://dx.doi.org/}%
\providecommand \selectlanguage [0]{\@gobble}%
\providecommand \bibinfo  [0]{\@secondoftwo}%
\providecommand \bibfield  [0]{\@secondoftwo}%
\providecommand \translation [1]{[#1]}%
\providecommand \BibitemOpen [0]{}%
\providecommand \bibitemStop [0]{}%
\providecommand \bibitemNoStop [0]{.\EOS\space}%
\providecommand \EOS [0]{\spacefactor3000\relax}%
\providecommand \BibitemShut  [1]{\csname bibitem#1\endcsname}%
\let\auto@bib@innerbib\@empty
\bibitem [{\citenamefont {Neubert}(1998)}]{Neubert:1997gu}%
  \BibitemOpen
  \bibfield  {author} {\bibinfo {author} {\bibfnamefont {M.}~\bibnamefont
  {Neubert}},\ }\href {\doibase 10.1142/9789812812667_0003} {\bibfield
  {journal} {\bibinfo  {journal} {Adv. Ser. Direct. High Energy Phys.}\
  }\textbf {\bibinfo {volume} {15}},\ \bibinfo {pages} {239} (\bibinfo {year}
  {1998})},\ \Eprint {http://arxiv.org/abs/hep-ph/9702375}
  {arXiv:hep-ph/9702375} \BibitemShut {NoStop}%
\bibitem [{\citenamefont {Uraltsev}(2001)}]{Uraltsev:2000qw}%
  \BibitemOpen
  \bibfield  {author} {\bibinfo {author} {\bibfnamefont {N.}~\bibnamefont
  {Uraltsev}},\ }in\ \href {\doibase 10.1142/9789812810458_0034} {\emph
  {\bibinfo {booktitle} {At the frontier of Particle Physics}}},\ \bibinfo
  {editor} {edited by\ \bibinfo {editor} {\bibfnamefont {M.}~\bibnamefont
  {Shifman}}\ and\ \bibinfo {editor} {\bibfnamefont {B.}~\bibnamefont
  {Ioffe}}}\ (\bibinfo {year} {2001})\ \Eprint
  {http://arxiv.org/abs/hep-ph/0010328} {arXiv:hep-ph/0010328} \BibitemShut
  {NoStop}%
\bibitem [{\citenamefont {Lenz}\ and\ \citenamefont
  {Rauh}(2013)}]{Lenz:2013aua}%
  \BibitemOpen
  \bibfield  {author} {\bibinfo {author} {\bibfnamefont {A.}~\bibnamefont
  {Lenz}}\ and\ \bibinfo {author} {\bibfnamefont {T.}~\bibnamefont {Rauh}},\
  }\href {\doibase 10.1103/PhysRevD.88.034004} {\bibfield  {journal} {\bibinfo
  {journal} {Phys. Rev. D}\ }\textbf {\bibinfo {volume} {88}},\ \bibinfo
  {pages} {034004} (\bibinfo {year} {2013})},\ \Eprint
  {http://arxiv.org/abs/1305.3588} {arXiv:1305.3588 [hep-ph]} \BibitemShut
  {NoStop}%
\bibitem [{\citenamefont {Lenz}(2015)}]{Lenz:2014jha}%
  \BibitemOpen
  \bibfield  {author} {\bibinfo {author} {\bibfnamefont {A.}~\bibnamefont
  {Lenz}},\ }\href {\doibase 10.1142/S0217751X15430058} {\bibfield  {journal}
  {\bibinfo  {journal} {Int. J. Mod. Phys. A}\ }\textbf {\bibinfo {volume}
  {30}},\ \bibinfo {pages} {1543005} (\bibinfo {year} {2015})},\ \Eprint
  {http://arxiv.org/abs/1405.3601} {arXiv:1405.3601 [hep-ph]} \BibitemShut
  {NoStop}%
\bibitem [{\citenamefont {Kirk}\ \emph {et~al.}(2017)\citenamefont {Kirk},
  \citenamefont {Lenz},\ and\ \citenamefont {Rauh}}]{Kirk:2017juj}%
  \BibitemOpen
  \bibfield  {author} {\bibinfo {author} {\bibfnamefont {M.}~\bibnamefont
  {Kirk}}, \bibinfo {author} {\bibfnamefont {A.}~\bibnamefont {Lenz}}, \ and\
  \bibinfo {author} {\bibfnamefont {T.}~\bibnamefont {Rauh}},\ }\href {\doibase
  10.1007/JHEP12(2017)068} {\bibfield  {journal} {\bibinfo  {journal} {JHEP}\
  }\textbf {\bibinfo {volume} {12}},\ \bibinfo {pages} {068} (\bibinfo {year}
  {2017})},\ \bibinfo {note} {[Erratum: JHEP {\bf 06} (2020) 162]},\ \Eprint
  {http://arxiv.org/abs/1711.02100} {arXiv:1711.02100 [hep-ph]} \BibitemShut
  {NoStop}%
\bibitem [{\citenamefont {Cheng}(2018)}]{Cheng:2018rkz}%
  \BibitemOpen
  \bibfield  {author} {\bibinfo {author} {\bibfnamefont {H.-Y.}\ \bibnamefont
  {Cheng}},\ }\href {\doibase 10.1007/JHEP11(2018)014} {\bibfield  {journal}
  {\bibinfo  {journal} {JHEP}\ }\textbf {\bibinfo {volume} {11}},\ \bibinfo
  {pages} {014} (\bibinfo {year} {2018})},\ \Eprint
  {http://arxiv.org/abs/1807.00916} {arXiv:1807.00916 [hep-ph]} \BibitemShut
  {NoStop}%
\bibitem [{\citenamefont {Gratrex}\ \emph {et~al.}(2022)\citenamefont
  {Gratrex}, \citenamefont {Melić},\ and\ \citenamefont
  {Nišandžić}}]{Gratrex:2022}%
  \BibitemOpen
  \bibfield  {author} {\bibinfo {author} {\bibfnamefont {J.}~\bibnamefont
  {Gratrex}}, \bibinfo {author} {\bibfnamefont {B.}~\bibnamefont {Melić}}, \
  and\ \bibinfo {author} {\bibfnamefont {I.}~\bibnamefont {Nišandžić}},\
  }\href@noop {} {\  (\bibinfo {year} {2022})},\ \Eprint
  {http://arxiv.org/abs/2204.11935} {arXiv:2204.11935 [physics.hep-ph]}
  \BibitemShut {NoStop}%
\bibitem [{\citenamefont {Aaij}\ \emph {et~al.}(2018)\citenamefont {Aaij} \emph
  {et~al.}}]{Aaij:2018dso}%
  \BibitemOpen
  \bibfield  {author} {\bibinfo {author} {\bibfnamefont {R.}~\bibnamefont
  {Aaij}} \emph {et~al.} (\bibinfo {collaboration} {LHCb Collaboration}),\
  }\href {\doibase 10.1103/PhysRevLett.121.092003} {\bibfield  {journal}
  {\bibinfo  {journal} {Phys. Rev. Lett.}\ }\textbf {\bibinfo {volume} {121}},\
  \bibinfo {pages} {092003} (\bibinfo {year} {2018})},\ \Eprint
  {http://arxiv.org/abs/1807.02024} {arXiv:1807.02024 [hep-ex]} \BibitemShut
  {NoStop}%
\bibitem [{\citenamefont {Aaij}\ \emph {et~al.}(2019)\citenamefont {Aaij} \emph
  {et~al.}}]{Aaij:2019lwg}%
  \BibitemOpen
  \bibfield  {author} {\bibinfo {author} {\bibfnamefont {R.}~\bibnamefont
  {Aaij}} \emph {et~al.} (\bibinfo {collaboration} {LHCb Collaboration}),\
  }\href {\doibase 10.1103/PhysRevD.100.032001} {\bibfield  {journal} {\bibinfo
   {journal} {Phys. Rev. D}\ }\textbf {\bibinfo {volume} {100}},\ \bibinfo
  {pages} {032001} (\bibinfo {year} {2019})},\ \Eprint
  {http://arxiv.org/abs/1906.08350} {arXiv:1906.08350 [hep-ex]} \BibitemShut
  {NoStop}%
\bibitem [{\citenamefont {Aaij}\ \emph {et~al.}(2022)\citenamefont {Aaij} \emph
  {et~al.}}]{Aaij:2021}%
  \BibitemOpen
  \bibfield  {author} {\bibinfo {author} {\bibfnamefont {R.}~\bibnamefont
  {Aaij}} \emph {et~al.} (\bibinfo {collaboration} {LHCb Collaboration}),\
  }\href {\doibase https://doi.org/10.1016/j.scib.2021.11.022} {\bibfield
  {journal} {\bibinfo  {journal} {Science Bulletin}\ }\textbf {\bibinfo
  {volume} {67}},\ \bibinfo {pages} {479} (\bibinfo {year} {2022})}\BibitemShut
  {NoStop}%
\bibitem [{\citenamefont {Zyla}\ \emph {et~al.}(2020)\citenamefont {Zyla} \emph
  {et~al.}}]{pdg}%
  \BibitemOpen
  \bibfield  {author} {\bibinfo {author} {\bibfnamefont {P.~A.}\ \bibnamefont
  {Zyla}} \emph {et~al.} (\bibinfo {collaboration} {Particle Data Group}),\
  }\href {\doibase 10.1093/ptep/ptaa104} {\bibfield  {journal} {\bibinfo
  {journal} {PTEP}\ }\textbf {\bibinfo {volume} {2020}},\ \bibinfo {pages}
  {083C01} (\bibinfo {year} {2020})}\BibitemShut {NoStop}%
\bibitem [{\citenamefont {Link}\ \emph {et~al.}(2002)\citenamefont {Link} \emph
  {et~al.}}]{Link:2002uhq}%
  \BibitemOpen
  \bibfield  {author} {\bibinfo {author} {\bibfnamefont {J.~M.}\ \bibnamefont
  {Link}} \emph {et~al.} (\bibinfo {collaboration} {FOCUS Collaboration}),\
  }\href {\doibase 10.1103/PhysRevLett.88.161801} {\bibfield  {journal}
  {\bibinfo  {journal} {Phys. Rev. Lett.}\ }\textbf {\bibinfo {volume} {88}},\
  \bibinfo {pages} {161801} (\bibinfo {year} {2002})},\ \Eprint
  {http://arxiv.org/abs/hep-ex/0202001} {arXiv:hep-ex/0202001} \BibitemShut
  {NoStop}%
\bibitem [{\citenamefont {Kushnirenko}\ \emph {et~al.}(2001)\citenamefont
  {Kushnirenko} \emph {et~al.}}]{Kushnirenko:2001}%
  \BibitemOpen
  \bibfield  {author} {\bibinfo {author} {\bibfnamefont {A.}~\bibnamefont
  {Kushnirenko}} \emph {et~al.} (\bibinfo {collaboration} {SELEX
  Collaboration}),\ }\href {\doibase 10.1103/PhysRevLett.86.5243} {\bibfield
  {journal} {\bibinfo  {journal} {Phys. Rev. Lett.}\ }\textbf {\bibinfo
  {volume} {86}},\ \bibinfo {pages} {5243} (\bibinfo {year} {2001})},\ \Eprint
  {http://arxiv.org/abs/hep-ex/0010014} {arXiv:hep-ex/0010014} \BibitemShut
  {NoStop}%
\bibitem [{\citenamefont {Mahmood}\ \emph {et~al.}(2001)\citenamefont {Mahmood}
  \emph {et~al.}}]{CLEO:2001}%
  \BibitemOpen
  \bibfield  {author} {\bibinfo {author} {\bibfnamefont {A.~H.}\ \bibnamefont
  {Mahmood}} \emph {et~al.} (\bibinfo {collaboration} {CLEO Collaboration}),\
  }\href {\doibase 10.1103/PhysRevLett.86.2232} {\bibfield  {journal} {\bibinfo
   {journal} {Phys. Rev. Lett.}\ }\textbf {\bibinfo {volume} {86}},\ \bibinfo
  {pages} {2232} (\bibinfo {year} {2001})},\ \Eprint
  {http://arxiv.org/abs/hep-ex/0011049} {arXiv:hep-ex/0011049} \BibitemShut
  {NoStop}%
\bibitem [{\citenamefont {Abudin\'en}\ \emph {et~al.}(2021)\citenamefont
  {Abudin\'en} \emph {et~al.}}]{Abudinen:2021}%
  \BibitemOpen
  \bibfield  {author} {\bibinfo {author} {\bibfnamefont {F.}~\bibnamefont
  {Abudin\'en}} \emph {et~al.} (\bibinfo {collaboration} {Belle II
  Collaboration}),\ }\href@noop {} {\bibfield  {journal} {\bibinfo  {journal}
  {Phys. Rev. Lett.}\ }\textbf {\bibinfo {volume} {127}},\ \bibinfo {pages}
  {211801} (\bibinfo {year} {2021})},\ \Eprint
  {http://arxiv.org/abs/2108.03216} {arXiv:2108.03216 [hep-ex]} \BibitemShut
  {NoStop}%
\bibitem [{\citenamefont {Abe}\ \emph {et~al.}(2010)\citenamefont {Abe} \emph
  {et~al.}}]{Abe:2010gxa}%
  \BibitemOpen
  \bibfield  {author} {\bibinfo {author} {\bibfnamefont {T.}~\bibnamefont
  {Abe}} \emph {et~al.} (\bibinfo {collaboration} {Belle II Collaboration}),\
  }\href@noop {} {\  (\bibinfo {year} {2010})},\ \Eprint
  {http://arxiv.org/abs/1011.0352} {arXiv:1011.0352 [physics.ins-det]}
  \BibitemShut {NoStop}%
\bibitem [{\citenamefont {Akai}\ \emph {et~al.}(2018)\citenamefont {Akai},
  \citenamefont {Furukawa},\ and\ \citenamefont {Koiso}}]{Akai:2018mbz}%
  \BibitemOpen
  \bibfield  {author} {\bibinfo {author} {\bibfnamefont {K.}~\bibnamefont
  {Akai}}, \bibinfo {author} {\bibfnamefont {K.}~\bibnamefont {Furukawa}}, \
  and\ \bibinfo {author} {\bibfnamefont {H.}~\bibnamefont {Koiso}},\ }\href
  {\doibase 10.1016/j.nima.2018.08.017} {\bibfield  {journal} {\bibinfo
  {journal} {Nucl. Instrum. Meth. A}\ }\textbf {\bibinfo {volume} {907}},\
  \bibinfo {pages} {188} (\bibinfo {year} {2018})},\ \Eprint
  {http://arxiv.org/abs/1809.01958} {arXiv:1809.01958 [physics.acc-ph]}
  \BibitemShut {NoStop}%
\bibitem [{\citenamefont {Jadach}\ \emph {et~al.}(2000)\citenamefont {Jadach},
  \citenamefont {Ward},\ and\ \citenamefont {Was}}]{Jadach:1999vf}%
  \BibitemOpen
  \bibfield  {author} {\bibinfo {author} {\bibfnamefont {S.}~\bibnamefont
  {Jadach}}, \bibinfo {author} {\bibfnamefont {B.~F.~L.}\ \bibnamefont {Ward}},
  \ and\ \bibinfo {author} {\bibfnamefont {Z.}~\bibnamefont {Was}},\ }\href
  {\doibase 10.1016/S0010-4655(00)00048-5} {\bibfield  {journal} {\bibinfo
  {journal} {Comput. Phys. Commun.}\ }\textbf {\bibinfo {volume} {130}},\
  \bibinfo {pages} {260} (\bibinfo {year} {2000})},\ \Eprint
  {http://arxiv.org/abs/hep-ph/9912214} {arXiv:hep-ph/9912214 [hep-ph]}
  \BibitemShut {NoStop}%
\bibitem [{\citenamefont {Sjöstrand}\ \emph {et~al.}(2015)\citenamefont
  {Sjöstrand} \emph {et~al.}}]{Sjostrand:2014zea}%
  \BibitemOpen
  \bibfield  {author} {\bibinfo {author} {\bibfnamefont {T.}~\bibnamefont
  {Sjöstrand}} \emph {et~al.},\ }\href {\doibase 10.1016/j.cpc.2015.01.024}
  {\bibfield  {journal} {\bibinfo  {journal} {Comput. Phys. Commun.}\ }\textbf
  {\bibinfo {volume} {191}},\ \bibinfo {pages} {159} (\bibinfo {year}
  {2015})},\ \Eprint {http://arxiv.org/abs/1410.3012} {arXiv:1410.3012
  [hep-ph]} \BibitemShut {NoStop}%
\bibitem [{\citenamefont {Lange}(2001)}]{Lange:2001uf}%
  \BibitemOpen
  \bibfield  {author} {\bibinfo {author} {\bibfnamefont {D.~J.}\ \bibnamefont
  {Lange}},\ }\bibfield  {booktitle} {\emph {\bibinfo {booktitle}
  {{Proceedings, 7th International Conference on B physics at hadron machines
  (BEAUTY 2000): Maagan, Israel, September 13-18, 2000}}},\ }\href {\doibase
  10.1016/S0168-9002(01)00089-4} {\bibfield  {journal} {\bibinfo  {journal}
  {Nucl. Instrum. Meth. A}\ }\textbf {\bibinfo {volume} {462}},\ \bibinfo
  {pages} {152} (\bibinfo {year} {2001})}\BibitemShut {NoStop}%
\bibitem [{\citenamefont {Agostinelli}\ \emph {et~al.}(2003)\citenamefont
  {Agostinelli} \emph {et~al.}}]{Agostinelli:2002hh}%
  \BibitemOpen
  \bibfield  {author} {\bibinfo {author} {\bibfnamefont {S.}~\bibnamefont
  {Agostinelli}} \emph {et~al.} (\bibinfo {collaboration} {GEANT4
  Collaboration}),\ }\href {\doibase 10.1016/S0168-9002(03)01368-8} {\bibfield
  {journal} {\bibinfo  {journal} {Nucl. Instrum. Meth. A}\ }\textbf {\bibinfo
  {volume} {506}},\ \bibinfo {pages} {250} (\bibinfo {year}
  {2003})}\BibitemShut {NoStop}%
\bibitem [{\citenamefont {Kuhr}\ \emph {et~al.}(2019)\citenamefont {Kuhr},
  \citenamefont {Pulvermacher}, \citenamefont {Ritter}, \citenamefont {Hauth},\
  and\ \citenamefont {Braun}}]{Kuhr:2018lps}%
  \BibitemOpen
  \bibfield  {author} {\bibinfo {author} {\bibfnamefont {T.}~\bibnamefont
  {Kuhr}}, \bibinfo {author} {\bibfnamefont {C.}~\bibnamefont {Pulvermacher}},
  \bibinfo {author} {\bibfnamefont {M.}~\bibnamefont {Ritter}}, \bibinfo
  {author} {\bibfnamefont {T.}~\bibnamefont {Hauth}}, \ and\ \bibinfo {author}
  {\bibfnamefont {N.}~\bibnamefont {Braun}} (\bibinfo {collaboration} {Belle II
  framework software group}),\ }\href {\doibase 10.1007/s41781-018-0017-9}
  {\bibfield  {journal} {\bibinfo  {journal} {Comput. Softw. Big Sci.}\
  }\textbf {\bibinfo {volume} {3}},\ \bibinfo {pages} {1} (\bibinfo {year}
  {2019})},\ \Eprint {http://arxiv.org/abs/1809.04299} {arXiv:1809.04299
  [physics.comp-ph]} \BibitemShut {NoStop}%
\bibitem [{\citenamefont {Bertacchi}\ \emph {et~al.}(2021)\citenamefont
  {Bertacchi} \emph {et~al.}}]{BelleIITrackingGroup:2020hpx}%
  \BibitemOpen
  \bibfield  {author} {\bibinfo {author} {\bibfnamefont {V.}~\bibnamefont
  {Bertacchi}} \emph {et~al.} (\bibinfo {collaboration} {Belle II tracking
  group}),\ }\href {\doibase 10.1016/j.cpc.2020.107610} {\bibfield  {journal}
  {\bibinfo  {journal} {Comput. Phys. Commun.}\ }\textbf {\bibinfo {volume}
  {259}},\ \bibinfo {pages} {107610} (\bibinfo {year} {2021})},\ \Eprint
  {http://arxiv.org/abs/2003.12466} {arXiv:2003.12466 [physics.ins-det]}
  \BibitemShut {NoStop}%
\bibitem [{\citenamefont {Kurz}()}]{Simon:1960}%
  \BibitemOpen
  \bibfield  {author} {\bibinfo {author} {\bibfnamefont {S.}~\bibnamefont
  {Kurz}},\ }\bibfield  {booktitle} {\emph {\bibinfo {booktitle} {{Proceedings,
  Connecting the Dots 2020}}},\ }\href@noop {} {\ }\bibinfo {note}
  {PROC-CTD2020-40,
  \href{https://docs.belle2.org/record/1960}{BELLE2-CONF-PROC-2020-009},
  \href{https://doi.org/10.5281/zenodo.4088760}}\BibitemShut {NoStop}%
\bibitem [{\citenamefont {Krohn}\ \emph {et~al.}(2020)\citenamefont {Krohn}
  \emph {et~al.}}]{treefitter}%
  \BibitemOpen
  \bibfield  {author} {\bibinfo {author} {\bibfnamefont {J.-F.}\ \bibnamefont
  {Krohn}} \emph {et~al.} (\bibinfo {collaboration} {Belle II analysis software
  group}),\ }\href {\doibase 10.1016/j.nima.2020.164269} {\bibfield  {journal}
  {\bibinfo  {journal} {Nucl. Instrum. Meth. A}\ }\textbf {\bibinfo {volume}
  {976}},\ \bibinfo {pages} {164269} (\bibinfo {year} {2020})},\ \Eprint
  {http://arxiv.org/abs/1901.11198} {arXiv:1901.11198 [hep-ex]} \BibitemShut
  {NoStop}%
\bibitem [{\citenamefont {Lesiak}\ \emph {et~al.}(2005)\citenamefont {Lesiak}
  \emph {et~al.}}]{bxic}%
  \BibitemOpen
  \bibfield  {author} {\bibinfo {author} {\bibfnamefont {T.}~\bibnamefont
  {Lesiak}} \emph {et~al.} (\bibinfo {collaboration} {Belle Collaboration}),\
  }\href {\doibase https://doi.org/10.1016/j.physletb.2004.11.038} {\bibfield
  {journal} {\bibinfo  {journal} {Phys. Lett. B}\ }\textbf {\bibinfo {volume}
  {605}},\ \bibinfo {pages} {237} (\bibinfo {year} {2005})}\BibitemShut
  {NoStop}%
\bibitem [{\citenamefont {Aaij}\ \emph {et~al.}(2020)\citenamefont {Aaij} \emph
  {et~al.}}]{xiclhcb}%
  \BibitemOpen
  \bibfield  {author} {\bibinfo {author} {\bibfnamefont {R.}~\bibnamefont
  {Aaij}} \emph {et~al.} (\bibinfo {collaboration} {LHCb Collaboration}),\
  }\href {\doibase 10.1103/PhysRevD.102.071101} {\bibfield  {journal} {\bibinfo
   {journal} {Phys. Rev. D}\ }\textbf {\bibinfo {volume} {102}},\ \bibinfo
  {pages} {071101} (\bibinfo {year} {2020})}\BibitemShut {NoStop}%
\bibitem [{\citenamefont {Niu}\ \emph {et~al.}(2022)\citenamefont {Niu},
  \citenamefont {Wang},\ and\ \citenamefont {Zhao}}]{xicppred}%
  \BibitemOpen
  \bibfield  {author} {\bibinfo {author} {\bibfnamefont {P.-Y.}\ \bibnamefont
  {Niu}}, \bibinfo {author} {\bibfnamefont {Q.}~\bibnamefont {Wang}}, \ and\
  \bibinfo {author} {\bibfnamefont {Q.}~\bibnamefont {Zhao}},\ }\href@noop {}
  {\bibfield  {journal} {\bibinfo  {journal} {Phys. Lett. B}\ }\textbf
  {\bibinfo {volume} {826}},\ \bibinfo {pages} {136916} (\bibinfo {year}
  {2022})}\BibitemShut {NoStop}%
\bibitem [{\citenamefont {Johnson}(1949)}]{johnson}%
  \BibitemOpen
  \bibfield  {author} {\bibinfo {author} {\bibfnamefont {N.~L.}\ \bibnamefont
  {Johnson}},\ }\href {\doibase 10.1093/biomet/36.1-2.149} {\bibfield
  {journal} {\bibinfo  {journal} {Biometrika}\ }\textbf {\bibinfo {volume}
  {36}},\ \bibinfo {pages} {149} (\bibinfo {year} {1949})}\BibitemShut
  {NoStop}%
\bibitem [{\citenamefont {Bilka}\ \emph {et~al.}(2020)\citenamefont {Bilka}
  \emph {et~al.}}]{Bilka:2020kgr}%
  \BibitemOpen
  \bibfield  {author} {\bibinfo {author} {\bibfnamefont {T.}~\bibnamefont
  {Bilka}} \emph {et~al.},\ }\href {\doibase 10.1051/epjconf/202024502023}
  {\bibfield  {journal} {\bibinfo  {journal} {EPJ Web Conf.}\ }\textbf
  {\bibinfo {volume} {245}},\ \bibinfo {pages} {02023} (\bibinfo {year}
  {2020})}\BibitemShut {NoStop}%
\end{thebibliography}%

\end{document}